\documentclass[conference]{IEEEtran}
\makeatletter
\def\ps@headings{%
\def\@oddhead{\mbox{}\scriptsize\rightmark \hfil \thepage}%
\def\@evenhead{\scriptsize\thepage \hfil \leftmark\mbox{}}%
\def\@oddfoot{}%
\def\@evenfoot{}}
\makeatother
\IEEEoverridecommandlockouts
\usepackage{amsmath,amssymb,amsfonts}
\usepackage{graphicx}
\usepackage{textcomp}
\usepackage{xcolor}
\usepackage{url}

\usepackage{algorithm}
\usepackage{algpseudocode}
\usepackage{color,soul}

\begin{document}
\bstctlcite{IEEEexample:BSTcontrol}

\title{Privacy-Aware Agent Collaboration for Dynamic VR Slice Management in 6G SD-RAN}


\author{
\IEEEauthorblockN{Khaled M. Naguib \textsuperscript{\textsection, \textdagger, 1}, Soumaya Cherkaoui\textsuperscript{\textsection, 2}, Mahmoud M. Elmesalawy\textsuperscript{\textparagraph, 3}, Ahmed M. Abd El-Haleem\textsuperscript{\textparagraph, 4},\\ and Ibrahim I. Ibrahim\textsuperscript{\textparagraph, 5}}\\

\IEEEauthorblockA{\textsuperscript{\textdagger}CCAS Department, School of Engineering, NewGiza University (NGU), Giza, Egypt}
\IEEEauthorblockA{\textsuperscript{\textsection}Department of Computer and Software Engineering, Polytechnique Montréal, Montréal, Canada}
\IEEEauthorblockA{\textsuperscript{\textparagraph}Electronics and Communications, Helwan Univ.
Cairo, Egypt}
\IEEEauthorblockA{Email: \textsuperscript{1}khaled.abdelmoneim@polymtl.ca, \textsuperscript{2}soumaya.cherkaoui@polymtl.ca, \\ \textsuperscript{3}melmesalawy@h-eng.helwan.edu.eg, \textsuperscript{4}ahmed.abdelkhaliq@h-eng.helwan.edu.eg, \textsuperscript{5}iiibrahim1953@gmail.com}}

\maketitle

\begin{abstract}


Ultra-low latency and high throughput are required for Virtual Reality (VR) services in 6G networks, which presents critical challenges for Software-Defined Radio Access Networks (SD-RANs) dynamic resource management. This work propose a mobility-driven, privacy-aware Multi-Agent Reinforcement Learning (MARL) framework for VR slice management, in which cooperative agents maximize resource distribution over end-to-end VR links while protecting the privacy of user data. Our approach incorporates mobility prediction and an information bottleneck encoder to facilitate effective and secure agent collaboration. In simulations, comparisons with traditional methods are studied which show up to 34\% throughput improvement, 28\% fewer resources, and 85\% less privacy leakage, guaranteeing dependable immersive VR experiences in future 6G environments.

\end{abstract}


%
\IEEEpeerreviewmaketitle

\section{Introduction}

\IEEEPARstart{I}mmersive applications such as Virtual Reality (VR), Augmented Reality (AR), and the metaverse require ultra-low latency and high data rates that exceed current 5G capabilities \cite{9687530}. These demands make VR a key driver for 6G, where Ultra-Reliable Low-Latency Communications (URLLC) and enhanced Mobile Broadband (eMBB) services are expected to be supported via network slicing and AI-native control \cite{filali2023communication}. In this setting, Software-Defined Radio Access Networks (SD-RAN) aligned with Open Radio Access Network (O-RAN) provides the flexibility needed to instantiate and adapt slices \cite{abouaomar2022federated, 10433640}. However, efficient multi-slice resource management remains challenging under user mobility and time-varying traffic \cite{mlika2021network}. Multi-Agent Reinforcement Learning (MARL) offers a promising approach by enabling decentralized agents to learn cooperative resource allocation policies that meet VR Service Level Agreements (SLAs) while improving system throughput \cite{10399918}. Since coordination can reveal sensitive user and network information, privacy-preserving coordination (e.g., encoded message passing) can be integrated to enable collaboration without sharing raw observations \cite{lei_new_2023}.


In \cite{9606825,9430902}, envelope‑based analytical and heuristic models for VR delay/reliability prediction in 6G are validated via simulation and Markov queueing, showing accurate performance estimation under varying user loads. \cite{9999295} proposes Federated Deep RL for O‑RAN slicing, enabling decentralized, privacy‑preserving resource allocation with scalable 6G slicing. \cite{7047300,10059833} derive and verify an optimal throughput‑maximizing SD‑RAN policy, achieving at least 20\% throughput gain over conventional RAN, especially under heavy user loads and poor channel conditions. \cite{9749222} introduces an AI‑native slicing architecture for intelligent 6G slice coordination, while \cite{10472043} employs AI/ML to provide statistical QoS‑aware digital twins over 6G by integrating URLLC, sensing, and communications. \cite{10225766} applies MARL to Metaverse digital twins via the AAHC method, jointly optimizing computation offloading and channel allocation for low latency and high reliability. Finally, \cite{10313309,9509579} present PP‑MARL, combining homomorphic encryption, differential privacy, and split learning to protect shared information with low overhead and reduced bandwidth consumption.

Unlike prior work centered on SD-RAN resource allocation, throughput maximization, or VR delay modeling, this study jointly advances SD-RAN and VR by tackling dynamic resource management under user mobility. It addresses VR-specific ultra-low latency and high data rate E2E requirements using a privacy-aware MARL framework to optimize resource sharing within a dedicated VR slice. Key contributions include:
\begin{itemize}
\item A practical VR slice management model is designed for 6G SD‑RAN, capturing dynamic traffic patterns and user mobility.
\item Inter‑agent cooperation is enabled without exposing private user data via a cooperative MARL algorithm assisted by information‑bottleneck encoding.
\item The model jointly manages intra‑slice VR E2E links and inter‑slice (URLLC, eMBB, VR) links to satisfy SLAs and ensure fairness.
\item A comprehensive analysis of throughput, latency, resource usage, and privacy trade‑offs is provided, demonstrating substantial gains over DQN, NAF, and SD‑RAN NS baselines.
\end{itemize}
The paper is organized as follows: Section II presents the SD‑RAN slicing model and formulates the problem; Section III details the proposed solution; Section IV reports the experimental results; and Section V concludes the work.

\section{System Model and Problem Formulation}

\begin{figure}
\centering
\includegraphics[width=\columnwidth]{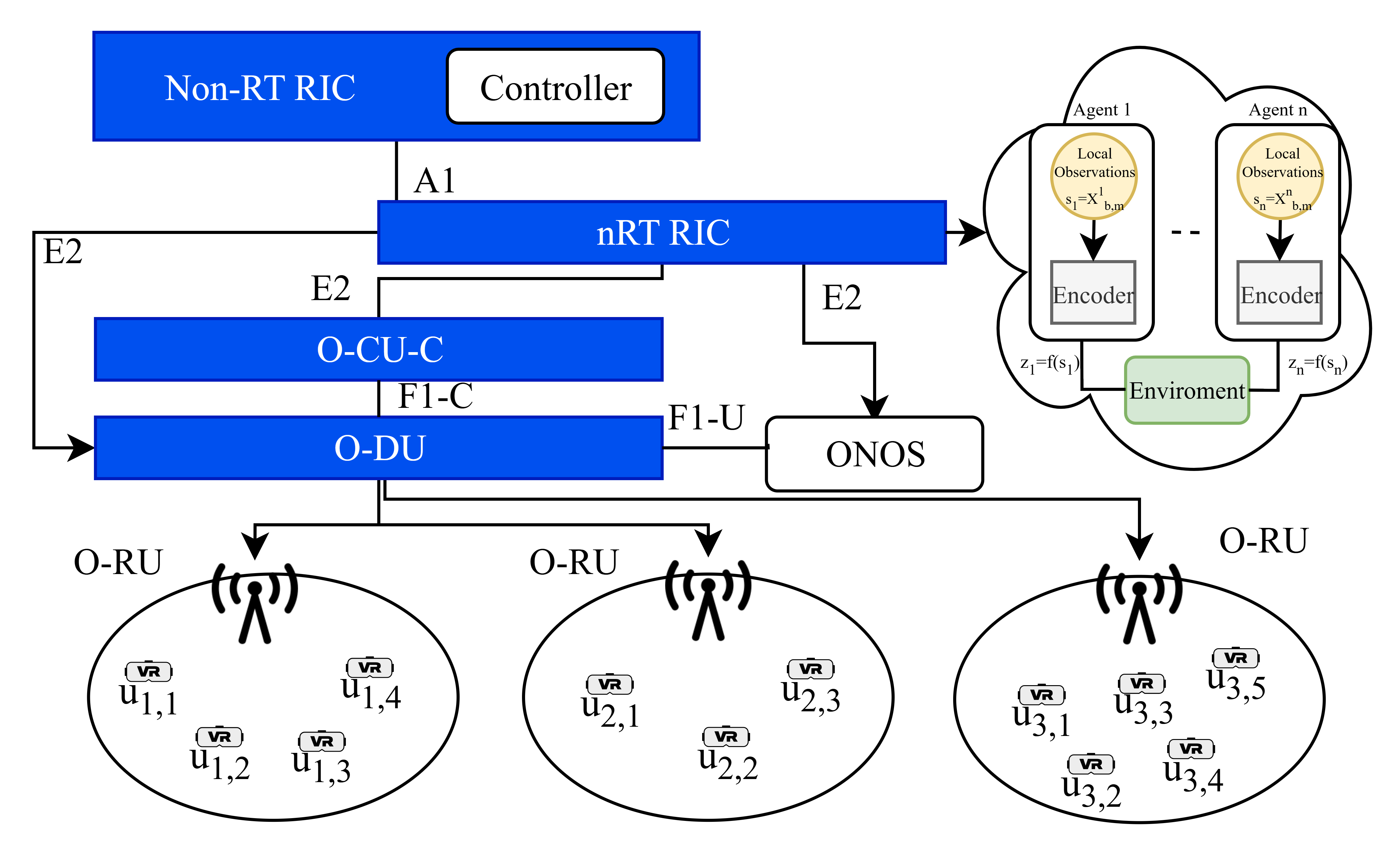} 
\caption{System Model}
\label{fig_system_model}
\end{figure}

This work considers a 6G system architecture based on SD-RAN, using Orthogonal Frequency Division Multiple Access (OFDMA) with $B$ O-RUs, O-DUs and O-CUs,  centrally located in each cell as shown in Fig. \ref{fig_system_model}. O-RUs are controlled by O-DUs and O-CUs so they are referred as Base Station (BS) in the rest of work. There are M User Equipments (UEs) homogeneously spread across the area and are engaged in diverse service virtual environment categories. Among them, VR traffic is extremely latency-sensitive and demands high data rates with E2E transmission guarantees.

In the system model, the radio interface operates over time-frequency resources, which are shared in the form of frames and further divided into subframes or Transmission Time Intervals (TTIs) of different length and subcarrier spacing. Various slices may use different numerologies over a single frame. The network dynamically assigns Resource Blocks (RBs) for each slice in each BS, where the VR slice would possess distinct uplink $\Theta_{b,m}^{UL}$ and downlink $\Theta_{b,m}^{DL}$ components to run within each BS. Each VR UE 
$m$ connected to BS $b$ is allocated RBs subsets denoted as $\theta_{b,n}^{UL} \subset \Theta_{b,n}^{UL}$ and $\theta_{b,n}^{DL} \subset \Theta_{b,n}^{DL}$ for uplink and downlink respectively.

Scheduling is driven by a control agent operating via the near-Real-Time RIC (nRT-RIC), which interacts with xApps via the E2 interface. The ONOS platform of SD-RAN is communicating with the controller through Non-Real Time RIC (Non-RT RIC), enabling programmable resource management and enforcing per-slice SLAs. Packet classification into its respective slice, real-time scheduling, and resource management, depending on network load, traffic dynamics, and channel quality, are performed for every incoming packet at the BS. A Gauss-Markov mobility model allows real-time user tracking and SLA satisfaction with seamless resource allocation. Slice-level utility functions also guide inter-slice and intra-slice resource allocation, exchanging channel conditions for service priorities. The system supports flexible and adaptive Round Robin (RB) scheduling across slices, enhancing the QoS for demanding services like VR while maintaining fairness across all services.

Define the total number of available RBs in the system as $N_s$. The RBs are distributed among different service slices such that eMBB, URLLC and VR slices. The resource allocation relationship can be expressed as:
\vspace{-0.3em}
\begin{equation}   
N_{s} = N_{eMBB} + N_{URLLC} + N_{VR}
\end{equation}

where $N_{eMBB}, N_{URLLC}$ and $N_{VR}$ are the number of resources assigned by the nRT-RIC according to users' mobility and available resources in the network eMBB, URLLC and VR slices, respectively. The uplink and downlink data rates achieved by a VR user $m \in M$ associated to base station $b \in B$ are expressed as:
\vspace{-0.3em}
\begin{equation}
R_{b,m}^{D} = BW \cdot \log_2\left(1 + \frac{p_{b,m}^{D} \cdot |h_{b,m}| \cdot d_{b,m}^{-\zeta}}{\sigma^2}\right), D \in \{UL,DL\}  
\end{equation}

where $BW$ is the channel bandwidth, $h_{b,m}$ represents the channel gain, $d_{b,m}$ is the distance between user $m$ and base station $b$, $\zeta$ is the path loss exponent,
$p_{b,m}^{D}$  are the uplink and downlink transmission powers, and 
$\sigma^2$  is the additive white Gaussian noise power variance. As such, the uplink and downlink transmission latency can be expressed as:
\vspace{-0.3em}

\begin{equation}
T_{b,m}^{D} = \frac{S_{m}^{D}}{R_{b,m}^{D}}  , D \in \{UL,DL\} 
\end{equation}

where $S_{m}^{DL}$ and $S_{m}^{UL}$ are the downlink and uplink packet sizes sent or received by user $m$, respectively.

Backhaul latency is also a significant factor in end-to-end delay, especially where centralized controllers are deployed in slice-level management. The total backhaul latency $T_{backhaul}$ has three components:
\vspace{-0.3em}
\begin{equation}
    T_{\text{backhaul}} = T_J + T_i + T_c
\end{equation}

where $T_J$ is the slice queue request time, $T_i$ is the processing time at SD-RAN network components, and $T_c$ represents the time taken by controller to make the decision. Most future technology in virtual environments involves interactions between two or more users, which comes with additional latency factors. Here, it is important to make the inter-user connection E2E with low latency so as not to interfere with the immersive and real-time aspects of the VR experience. However, due to the data exchange asymmetry, the amount of data sent or received by one user can be significantly smaller or larger than the data sent or received by the other. Therefore, the transmission latency for one user can differ from the reception latency for the other. The asymmetry requires both uplink and downlink latencies to be separated and balanced in the system model. Therefore, the E2E latency for a communication link between two users $i \in M$ and $j \in M$ is presented as:
\vspace{-0.3em}
\begin{equation}
    T^{total}_{i,j} = \alpha_{i,j} T_{b,i}^{UL} + T_{backhaul} + (1 - \alpha_{i,j}) T_{b,j}^{DL} 
\end{equation}

where $i,j \in M$, $\alpha \in [0,1]$ is a weighting factor used to scale the uplink and downlink latency terms so that both receiving and transmitting users' QoS requirements are met. Let $\beta_{b,m}\in\{0,1\}$ indicate whether BS $b$ serves VR user $m$, and $c_{i,j}\in\{0,1\}$ indicate whether an E2E VR link exists from $i$ to $j$. Hence, the overall data rate for the VR slice is expressed as:
\vspace{-0.3em}
\begin{equation}
R_{\text{total}} = \sum_{b=1}^{B} \sum_{i=1}^{M} \sum_{\substack{j=1 \\ j \ne i}}^{M} \Big[ 
c_{i,j}  \beta_{b,m} R_{b,m}^{D}  
+ \left(1 - c_{i,j} \right) \beta_{b,m} R_{b,m}^{D}
\Big]
\end{equation}

The optimization problem is aiming to maximize the total system throughput and can be expressed as:
\vspace{-0.3em}

\begin{align}
    \max_{\theta,\alpha} \quad & R_{\text{total}}  \\
    \text{s.t.} \quad 
    & T_{i,j}^{total} \leq T_{i,j}^{\text{th}}  \tag{i} \\
    & \alpha_{i,j} \in [0,1]  \tag{ii} \\
    & R_{b,m}^{DL} > R_{th}^{DL}  \hspace{1em} and \hspace{1em} R_{b,m}^{UL} > R_{th}^{UL}  \tag{iii} \\
    &  \sum_{m=1}^M \beta_{b,m} \theta_{b,m}^{D} \leq \Theta_{b,m}^{D}, D \in \{ UL,DL\}  \tag{iv} \\
    & \Theta_{b,m}^{UL} + \Theta_{b,m}^{DL} = N_{VR} \tag{v}
    \label{problem_formulation} \\
    &  I(X_{b,m}^{D}, z_{b,m}^{D}) \leq \epsilon^{D} \tag{vi}, D \in \{UL,DL\}
\end{align}

where $z_{b,m}^{UL}$ and $z_{b,m}^{DL}$ are the encoded messages shared for UL and DL coordination, respectively as well as $X_{b,m}^{DL}$ and $X_{b,m}^{UL}$ are the sensitive or private data that have not been shared among different SD-RAN components for uplink and downlink transmission, respectively, and $\epsilon^{DL}$ and $\epsilon^{UL}$ represent the maximum allowable privacy leakage levels in downlink and uplink transmissions, respectively. Smaller values indicate stronger privacy guarantees, and these values are determined using the differential privacy privacy budget concept.  In actuality, the Information Bottleneck encoder, which restricts the mutual information between raw states and shared representations, is used to enforce the value of $\epsilon$, which is determined by the intended trade-off between privacy and utility.  Due to asymmetric leakage risks caused by potential differences in user mobility and traffic patterns between the two directions, uplink and downlink privacy are enforced separately. Constraint (i) ensures that the total latency $T_{i,j}^{total}$ for communication between any two VR users is less than threshold $T_{i,j}^{\text{th}}$, as required to preserve the immersive quality of virtual worlds. Constraint (ii) enables flexible modeling of asymmetric traffic conditions when the latencies in the uplink and downlink differ due to unequal data sizes received or sent by the users. Constraints (iii) ensure that the data rates in the downlink and uplink directions, are above their target threshold values, thereby maintaining communication quality and preventing degradation in user experience. Constraints (iv) and (v) impose upper bounds on the overall assigned resource blocks to VR users served by a base station $b$, so that the total allocation is contained within the uplink and downlink capacity constraints. In addition, constraint (vi) ensure that sensitive data of user $m$ in BS $b$ is less than or equal to allowable privacy leakage. Hence, these constraints help the system minimize latency while adhering to user QoS requirements and resource constraints privately.

In our architecture, agents do not share the raw private variables $X^{UL}_{b,m}$ and $X^{DL}_{b,m}$; instead, each agent shares a compressed representation $z=f_{\phi}(s)$. Hence, privacy leakage is quantified by the mutual information between the private variables X and the shared messages z, i.e., $I(X;z)$, and the budgets $\epsilon^{UL},\epsilon^{DL}$ bound the maximum allowable leakage in each direction. This formulation is consistent with the information-bottleneck encoder in Section III, which learns z to preserve task-relevant information while discarding private information.

\section{The Proposed Solution}

An intelligent xAPP is deployed in the SD-RAN architecture using ONOS-based nRT-RIC to enable real-time slice migration and resource optimization. The xAPP leverages UE measurements and applies DRL with a defined state space (e.g., mobility, signal quality), action space (slice/resource decisions), and reward function to maximize network throughput. A mobility predictor enhances slice allocation by forecasting user movement and preemptively adapting resources. Network Simulator 3 (NS-3) \cite{ns3} and SUMO \cite{SUMO} simulate user mobility for training. The DRL model observes state $s_t = (x_t, y_t)$, where $x_t$ and $y_t$ is the position of user, selects action $a_t$, and updates policy based on reward $r_t$ computed via the Gauss-Markov model \cite{10150581657}. This enables mobility-aware, intra-slice resource optimization.

The mobility prediction using DRL for the SD-RAN slice optimization procedure is trained and utilized to NS-3 and SUMO mobility traces to learn real-world user behavior. It perceives the present state $s_t = (x_t, y_t)$ at every step, selects an action $a_t$ following the policy $\pi(a_t | s_t)$, moves to a new state $s_{t+1}$, and obtains a reward $r_t$ from the Gauss-Markov model. Transitions are stored and used for model training iteratively, converging to the optimal policy $\pi^*$.

In our proposed solution, MARL algorithm is used in which every agent controls one E2E VR connection. Agents run per BS, but they work together to optimize resource allocation across VR users. The goal is to calculate the optimal $\alpha_{i,j}$ per connection to maximize attainable throughput while guaranteeing fair resource allocation across the users of the same BS.


Every agent acts upon its environment according to:
\begin{itemize}
    \item State ($s_{i,t}$): contains active VR users over BS, user QoS requirements (target rate and delay), Channel Quality Information (CQI), and current resource utilization which is represented earlier in as $X_{n,m}^{UL}$ and $X_{n,m}^{DL}$.

    \item Action ($a_{i,t}$): specifies the value of $\alpha_{i,j}$ for each VR E2E link and number of resources needed for each link.
    
    \item Reward ($r_{i,t}$): gives reward for higher achievable rates and penalizes skewness in allocation among users, given by:
    \begin{equation}
        r_{i,t} = R_{total} - \lambda \cdot Penalty^D , D \in \{UL, DL\} \label{9}
    \end{equation}

\end{itemize}

where $\lambda$ is a tunable weight for penalizing violations. To ensure reproducibility, the penalty term in \eqref{9} is explicitly defined using the positive-part operator $[x]^+=\max(0,x)$. Let $\mathcal{L}$ denote the set of active VR E2E links and let $r_\ell$ be the achieved rate of link $\ell\in\mathcal{L}$ with mean $\bar r=\frac{1}{|\mathcal{L}|}\sum_{\ell\in\mathcal{L}} r_\ell$. We set
\begin{align}
\mathrm{Penalty^D} \;=\;& \lambda_T \sum_{(i,j)\in\mathcal{L}} \big[T^{\mathrm{total}}_{i,j}-T^{\mathrm{th}}_{i,j}\big]^+ \nonumber
\\&+ \lambda_R \sum_{b}\sum_{m} \Big(\big[R^{D}_{\mathrm{th}}-R^{D}_{b,m}\big]^+ + \nonumber 
\\&+ \lambda_{RB}\sum_{b}\Big[\sum_{m}\beta_{b,m}\theta^{D}_{b,m}-\Theta^{D}_{b}\Big]^+ \nonumber
\\&+ \lambda_P \sum_{b}\sum_{m}\Big(\big[I(X^{D}_{b,m};z^{D}_{b,m})-\epsilon^{D}\big]^+ 
\end{align}

where $\lambda_T,\lambda_R,\lambda_{RB},\lambda_P\ge 0$ are tunable weights for latency, rate, RB-budget, and privacy-violation penalties, respectively. However, enabling cooperation among such MARL agents is generally accomplished by transmitting local observations or policy information, and this could accidentally expose sensitive or private user information such as mobility, traffic behavior, or service types. A privacy-preserving coordination scheme based on the Information Bottleneck (IB) paradigm is proposed in order to prevent this risk. In this setting, each MARL agent employs a learned encoder $f_\phi(s_{i,t})$ parameterized by neural network weights $\phi$, to map its local observation state $s_{i,t}$ to a compressed message $z_i = f_\phi(s_{i,t})$ before it is shared with centralized decision unit. The encoder is learned to maximize the mutual information between $z_i$ and the agent's policy-relevant features (i.e., useful information for resource coordination), while minimizing the information $s_{i,t}$ comprises about private variables. Formally, the encoder objective follows IB and formulated as: 

\begin{equation}
    \mathcal{L}_{IB} = \mathcal{L}_{RL}(a_{i,t},s_{i,t}) + \beta \cdot I(s_{i,t};z_i)
\end{equation}

where $\mathcal{L}_{RL}(a_{i,t},s_{i,t})$ is the agent loss function at state $s_t$ with an action $a_{i,t}$, $\beta$ is the private accuracy tradeoff hyperparameter and $I(s_{i,t};z_i)$ is the mutual information between compromised information and observations where it is approximated using the variational upper bound method \cite{9154315}. The framework of privacy aware agents MARL is illustrated in Fig \ref{fig_system_model}. Using synthetic mobility traces (SUMO + NS-3) and an auxiliary loss that strikes a balance between privacy leakage and policy-relevant utility, the encoder is pretrained offline. Since only forward passes through a lightweight neural network are needed, the encoder parameters are fixed during online deployment, guaranteeing minimal latency overhead. The encoder's runtime in NS-3 was profiled to determine latency, and it was continuously less than 0.1 ms per inference, indicating that it does not significantly increase the system's computational load.

In the MARL-based slice allocation algorithm in SD-RAN, every agent observes local state parameters—number of active VR users $U_i$, Channel Quality Indicators (CQIs) ${CQI}_u$, QoS demands $R_u, d_u$, and present loads and computes the best resource fraction $\alpha_{i,j}$ for the E2E link. Actions are evaluated by a reward function trading data rate maximization against fairness. Transitions are accumulated and employed to repeatedly update policies. The decentralized structure allows it to scale and adjust over VR links in real-time.

At each decision epoch, the SD‑RAN controller allocates UL/DL RB budgets per BS to the VR slice, and each BS hosts a cooperative MARL agent per active VR E2E link. Agent i observes a local state $s_{i,t}$ containing mobility, channel quality, QoS targets, and RB utilization, and selects an action $a_{i,t}$ that adjusts scheduling variables (e.g., UL/DL weighting factor $\alpha_{i,j}$ and RB allocation) to maximize long‑term throughput under SLA constraints. To preserve privacy, agents exchange neither raw observations nor measurements; instead, each applies an information‑bottleneck encoder to produce a compressed message $z_i$, which is sent to a coordination unit (e.g., nRT‑RIC/xApp). The unit aggregates these encoded messages to enhance multi‑link consistency within the slice and enforce RB‑budget constraints across slices, then returns coordination decisions to the agents.

\section{Results and Discussion}

To evaluate the proposed model, simulations were run and compared to different Reinforcement Learning (RL)-based slicing strategies. The simulation environment illustrates how RBs are distributed over different RAN slices, each optimized to support a specific service eMBB, URLLC, or VR. Each slice operates according to its own traffic pattern, Service Level Agreement (SLA) and mobility users' patterns where SLA specification is thus a key part of the resource management policy in network slicing.

Users were uniformly distributed in 600 m × 150 m. Transmission powers were 7 dBm for uplink and 23 dBm for downlink matches small-cell BS transmission limits reported in 3GPP TR 38.901 \cite{3gppTR38901}. The system bandwidth of 1 MHz was selected to balance computational load in NS-3 while preserving comparative validity, consistent with prior VR slicing works and the noise power density of -174 dBm follows standard thermal noise assumptions. Uplink and downlink transmission file sizes were altered randomly between 1 MB and 20 MB which capture realistic VR frame content variability, including compressed panoramic frames and 3D textures. The VR slice topology is dynamic according to the number of active VR users, where communication is modeled as VR-to-VR links with one user transmitting and another receiving. The simulation scenarios are focusing on the VR intra-slicing affected by the pre-assigned RBs for each slice by the inter-slicing algorithm. Radio resources were allocated with $TTI = 0.5 ms$ and subcarrier of 30 MHz with resource allocation decision every 25 ms (equivalent to 50 sub-frames) reflect practical control-plane update intervals in SD-RAN implementations. The performance of the proposed model is compared with two conventional RL  network slicing based algorithms: Deep Q-Network (DQN) method \cite{filali2023communication, 11178232} and Normalized Advantage Function (NAF) technique from \cite{8736846} . Additionally, another Network Slicing (NS) technique applied with SD-RAN scenario in \cite{10.1145/2999572.2999599} is included knowing that algorithm didn't account user mobility or VR link dynamics inclusion. Each MARL agent is implemented as a neural network with two fully connected ReLU layers, trained via Adam ($10^{-3}$ learning rate, discount factor 0.99). Experience replay with a buffer of 1000 transitions per agent and minibatches of size 64 is used, alongside an $\epsilon$-greedy exploration policy whose $\epsilon$ decays from 0.9 to 0.05 over 4000 steps. The mobility predictor is trained offline on SUMO‑derived traces integrated with NS‑3 simulations, while the information‑bottleneck encoder uses a lightweight network trained with a variational mutual‑information upper bound. Training stops when the moving‑average reward stabilizes over 200 consecutive episodes, and all experiments are averaged over multiple random seeds to ensure robustness.

\begin{figure}
\centering
\includegraphics[width=\columnwidth]{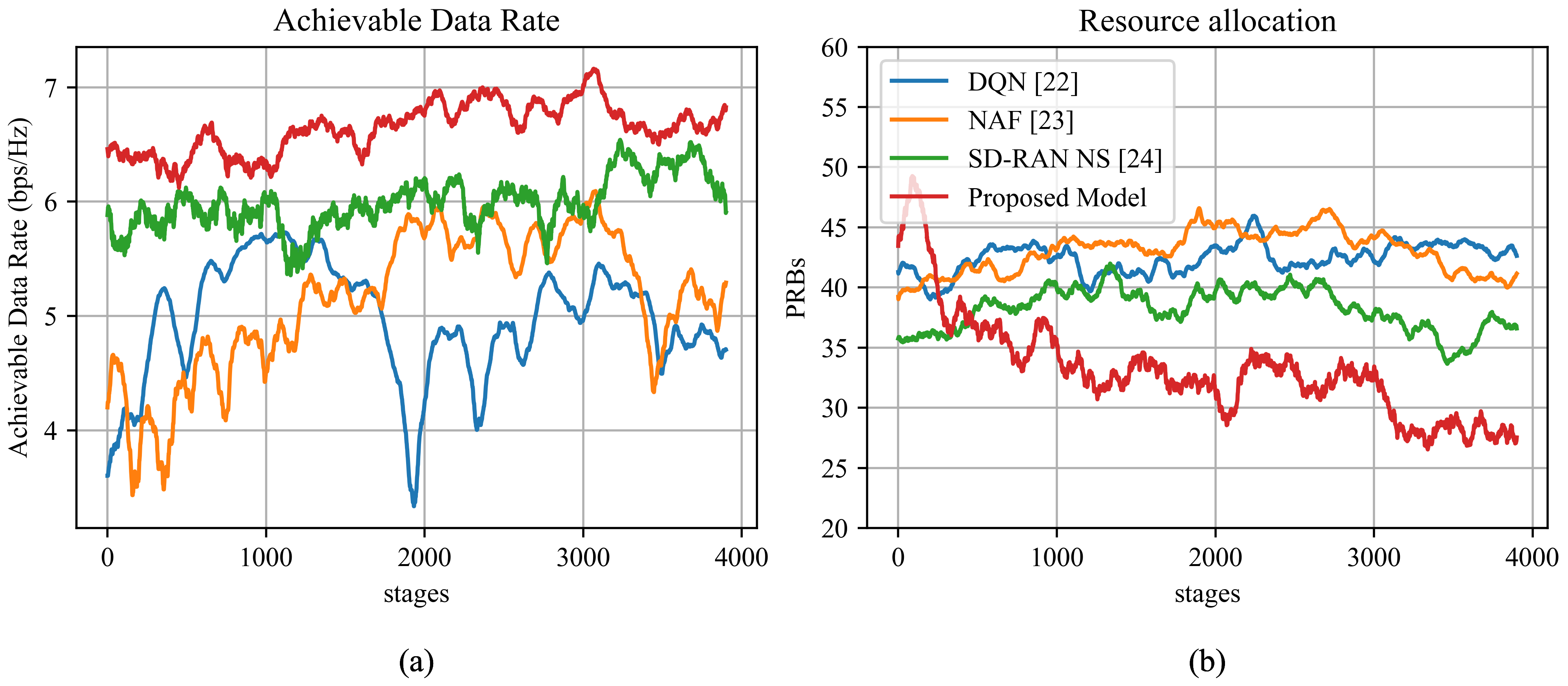} 
\caption{Achievable performance of the proposed MARL-based SD-RAN framework. (a) Average data rates (Mbps) during decision stages. (b) Resource block allocation patterns per stage (number of RBs).}
\label{fig_stages}
\end{figure}

Fig. \ref{fig_stages}(a) shows the data rate achieved during the decision stages of the simulation. It is clear that the mobility predictor and the $\alpha_{i,j}$ optimization in our MARL framework are largely responsible for the better data rate performance.  Our model proactively modifies allocations based on anticipated user movement, in contrast to baseline approaches that assign RBs statically or reactively.  This keeps resources from being underutilized. This improvement is because of two primary improvements: dynamic estimation of the number of RBs required per slice based on user mobility patterns, and optimal calculation of the weight factor $\alpha_{i,j}$ for each E2E VR link so that more optimal resource provisioning is supported. The reward function's fairness penalty, which deters unequal spectrum usage across VR links, is the direct cause of the more balanced RB allocation patterns seen in Fig. 3(b), which further supported.  Therefore, the design of the mobility-aware reward and $\alpha_{i,j}$ decision mechanism is intrinsically linked to both throughput and spectral efficiency gains. Moreover,our model's privacy-preserving mechanism serves a positive purpose by allowing users of various distributed base stations to share link-associated information in coded form. In addition to allowing effective RB allocation, this boosts user privacy among interconnected network segments.

\begin{figure}
\centering
\includegraphics[width=\columnwidth]{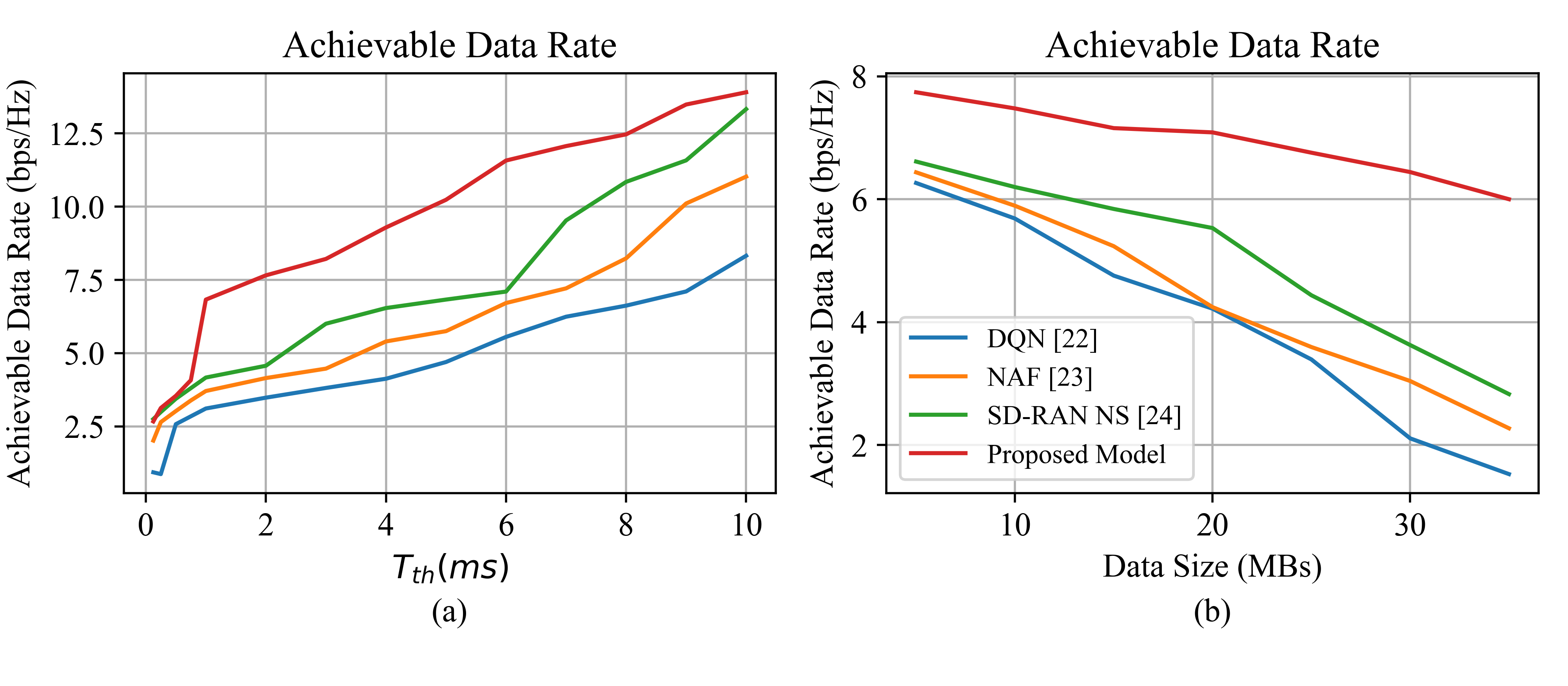} 
\caption{Impact of system parameters on achievable data rates. (a) Data rate (Mbps) vs. latency threshold $T_{th}$ (ms) of VR E2E links. (b) Data rate vs. transmitted file size (MB).}
\label{fig_data_rate}
\end{figure}

Fig. \ref{fig_data_rate}(a) illustrates the relation between the data rate that can be achieved and the time threshold of E2E VR links. The results show that with a larger time threshold, the system is more able to provide a higher data rate in all models. Especially, the proposed model has a much steeper increase in performance than baseline methods. This is primarily due to its effective resource allocation based on the unique requirements of each VR user in E2E link, derived from the global view of the SD-RAN framework. Once the threshold crosses 1 ms, the system stabilizes and operates even more seamlessly, enabling our model to surpass other methods by a steady margin. In addition, note that the incorporation of the privacy-preserving MARL mechanism incurs no extra backhaul latency. The utilized encoder simply converts the data representation without introducing computational overhead, since it is pretrained and optimized to be compatible with the system's latency constraint.

Figure \ref{fig_data_rate}(b) also plots achievable data rates against the file size being transmitted. As the file size increases, all the models experience some level of congestion, which influences throughput. However, the proposed model continues to exhibit superior performance, which is a result of its dynamic and distributed resource management strategy, which allows underutilized BSs to dynamically redistribute free spectrum to users with higher transmission demands. Consequently, even under high loads—i.e., when file sizes are 20 MB—our model still offers a much higher data rate than the other approaches, demonstrating its applicability to real-world VR environments with bandwidth intensity.

\begin{table}[htbp]
\centering
\caption{Comparison of privacy-preserving efficiency across different approaches, including slicing RL / FedRL baselines.}
\resizebox{\columnwidth}{!}{%
\begin{tabular}{|l|c|c|c|c|c|}
\hline
\textbf{Metric} & \textbf{DQN \cite{s22083031}} & \textbf{NAF \cite{8736846}} & \textbf{SD-RAN NS \cite{10.1145/2999572.2999599}} & \textbf{Slicing \cite{9852968}} & \textbf{Proposed Model} \\
\hline
Privacy Leakage Risk (0–1) & 0.82 & 0.76 & 0.56 & 0.48 & 0.12 \\
Shared Data Volume (KB/msg) & 150 & 130 & 95 & 120 & 40 \\
Performance Drop from Non-privacy (\%) & -- & -- & $\leq 6$ & $\leq 8$ & $\leq 2$ \\
Obfuscation / Encryption Overhead (\%) & -- & -- & 4.1 & 5.8 & 3.4 \\
Communication Overhead (Mbps) & -- & -- & 1.2 & 2.3 & 0.9 \\
Convergence Speed (Episodes to Stability) & 900 & 850 & 700 & 1000 & 650 \\
\hline
\end{tabular}
}
\label{tab:privacy_comparison_extended}
\end{table}

Table~\ref{tab:privacy_comparison_extended} evaluates privacy‑preserving efficiency in terms of privacy leakage risk, shared data volume, performance drop from non‑privacy baselines, and obfuscation overhead. Privacy leakage is quantified by the normalized probability of correctly inferring private user attributes (e.g., mobility and service type) from inter‑agent messages, estimated via an auxiliary classifier trained on exchanged encodings \cite{10313309}; the information‑bottleneck coefficient $\beta$ controls the privacy–utility trade‑off. Increasing $\beta$ intensifies compression, generally reducing leakage but potentially removing task‑relevant features, so a sensitivity study sweeps $\beta$ from 0.01 to 1 and records VR throughput and leakage risk. The proposed method achieves significantly lower leakage and minimal shared data volume (40 KB per message), indicating compact, efficient encoding. Compared to SD‑RAN NS , which suffers up to a 6\% performance drop, our framework limits degradation to about 2\%, while maintaining negligible obfuscation overhead, thus delivering an efficient privacy–reward trade‑off.

The proposed framework aligns with O-RAN by integrating nRT‑RIC xApps for slice-level resource optimization, with MARL agents using only locally available measurements (e.g., CQI, QoS, and mobility) already present in SD‑RAN control loops. Inter‑agent coordination overhead is reduced via information‑bottleneck encoding, compressing messages while preserving policy‑relevant features; encoder inference latency is below 0.1 ms per cycle, negligible compared to SD‑RAN update intervals. Though currently evaluated in single‑cell simulations, the decentralized MARL design naturally extends to multi‑cell deployments; practical challenges include controller synchronization, edge integration, and heterogeneous hardware constraints. The modular SD‑RAN architecture and lightweight privacy‑aware coordination make the framework promising for emerging 6G infrastructures.

\section{Conclusion}


This work proposed a privacy-aware MARL framework for VR slice management in 6G SD-RANs by addressing mobility dynamics and inter-agent privacy.  In comparison to traditional RL baselines, the model achieves significant improvements in throughput, latency, and privacy by combining mobility prediction with an information bottleneck encoder. The proposed model can be used for latency-critical services like AR, telemedicine, metaverse interactions, and collaborative robotics in addition to VR. The proposed work outperforms DQN, NAF, and SD-RAN NS in throughput (up to 34\% improvement), efficiency (up to 28\% fewer resources), and privacy (up to 85\% fewer leakages), with reduced data sharing and overhead. However, only simulated mobility traces and a small set of baselines are used in the current evaluation.  Future research will look into robustness against scalability to multi-cell scenarios, integration with edge computing, and real-world testbed deployment. Additionally, future work will assess its resilience to stronger privacy-inference adversaries (e.g., model inversion and attribute inference) under realistic inter-controller coordination.

\end{document}